\documentclass{aa}
\usepackage{astron}

\begin{document}
\thesaurus{06(06.13.1;03.20.5;02.18.7;02.16.2;03.13.4)}
\title{Integration of the radiative transfer equation for polarized
light: The exponential solution}
\author{M. Semel \and A. L{\'o}pez Ariste}
\institute{DASOP, URA 326, Observatoire de Paris, Section de Meudon}
\date{Received    ; Accepted    }
\titlerunning{The exponential solution}
\maketitle

\offprints{ M. Semel}
\mail{Meir.Semel@obspm.fr}
\begin{abstract}
The radiative transfer equation (RTE) for polarized light accepts a 
convenient exponential solution when the absorption matrix commutes with its 
integral. We characterize some of the matrix depth variations which are 
compatible with the  commutation condition. Eventually the vector solution 
may be diagonalized and one may obtain
four independent scalar solutions with four optical depths, complex in
general. When the commutation condition is not satisfied, one must resort
to a determination of an appropriate evolution operator, which is shown to be 
well determined mathematically, but whose
explicit form is, in general, not easy to apply in a numerical code.
However, we propose here an approach to solve a general
case not satisfying the commutation condition.
\end{abstract}

\section{Introduction}

The use of spectropolarimetry and Zeeman effect  to measure  
magnetic fields in the sun and stars, requires a transfer theory for 
polarized light,
in the presence of a magnetic field. Such transfer  equation 
was first written in the pioneering paper by Unno \cite*{Unn56}, where 
the effect of anomalous dispersion was not considered yet. Rachkowsky 
\cite*{Rac62,Rac67} was the first to include it, and obtained
a general  transfer  equation for polarized light,
\begin{equation}
\frac{d}{dz}\vec{I}=-\tens{K}\vec{I}+\vec{J} ,
\label{RTE}
\end{equation}
in terms of the Stokes parameters I, Q, U and V presented
as the components of vector $\vec{I}$. 
This equation {\it resembles} the transfer equation for the 
scalar case when polarization is ignored.
The  scalar intensity is replaced by the Stokes vector $ \vec{I}$.
The emision term  is now  a four component
 vector $\vec{J}$, and the scalar absorption coefficient 
becomes a $4\times 4$ matrix $\tens{K}$ which describes absorption 
(including anomalous dispersion) 
in the presence of Zeeman effect. The variable $z$ parameterizes the light 
path.
The transformation from a single equation to a system of four equations
(abridged in a vectorial form) 
changes drastically the nature of the problem and no general explicit
analytical solution has been proposed so far.

 The first particular solution was obtained by Unno, who applied the equation 
to the case of a Milne-Eddington atmosphere. Rachkowsky \cite*{Rac67}, after 
including anomalous dispersion, solved the RTE for the same homogeneous 
atmosphere. 

 Several procedures were successful in solving the equation numerically
\nocite{BS69,Wit74,RM89,LD76}
(e.g. Beckers \& Schr{\"o}ter 1969, Wittman 1974, Rees, Murphy \& Durrant 1989, 
Landi  Degl'Innocenti 1976) even for the general case, by subdividing the 
atmosphere into 
numerous layers. The lasts are chosen optically thin in the upper atmosphere 
and eventually 
an Unno--Rachkowsky solution is taken for the deepest layer, down to optical 
depth infinity.
 Difficulties were encountered as well, depending on the 
particular method chosen to solve the radiative transfer in each individual 
sublayer. An universal
technique, like the Runge--Kutta method, may not be the best approach for a 
particular case, mainly because of different scales of variation. The 
mathematical justifications 
are often not rigourous  and numerical tests  are always necessary.

The advantage of an eventual analytical solution is obvious.  But, up to now, 
they have been always restricted to
 homogeneous atmospheric  models where only the LTE source function was depth 
dependent. The constant $\tens{K}$ matrix has been handled with 
different mathematical techniques: for instance Kjeldseth Moe \cite*{KM68} 
and Stenflo\cite*{Ste71,Ste94} used $\tens{K}$-diagonalization,
van Ballegooijen\cite*{VB85} preferred Jones calculus. 

Now, for the solar case, variations with depth of both the thermodynamical 
parameters describing the at\-mos\-phe\-re and the magnetic field, are not 
negligible. For instance, Ruiz Cobo \& del Toro Iniesta\cite*{RCdTI92,RCdTI94},
 and del Toro Iniesta \& Ruiz Cobo \cite*{dTIRC96b,dTIRC96a}
 using numerical inversion of the observed 
Stokes profiles have confirmed the need for inhomogeneous models
\nocite{CMPRC+94,WPdTIRC+97a,WPdTIRC+97b,WPdTIRC+97c,dTIRC96a}
(see Collados {\em et al.} 1994,
Westendorp Plaza {\em et al.} 1997a,1997b,1997c, and 
see also del Toro Iniesta \& Ruiz Cobo 1996a for a review).
All these works raise the interest in analytical methods dealing with 
non-homogeneous atmospheric models, i.e. with non-constant $\tens{K}$ matrices.

At  first sight, the RTE for polarized light does not seem more complicated to
solve than its scalar equivalent (RTE for pure light intensity, with no 
polarisation):
$$\frac{d}{dz}I=-\kappa I+J,$$
where as usual, $\kappa$ is the absorption coefficient, $I$ and $J$ are the 
scalar intensity and emission function respectively and $z$ the geometrical 
path. This equation has an explicit formal solution, for the layer 
$z_0< t < z$ :
\begin{equation}
I (z) = \int_{z_0}^z e^{-\int_t^z \kappa dt'} J(t) dt +  
e^{-\int_{z_0}^z \kappa dt'} I(z_0).
\label{scalar}
\end{equation}
A direct extrapolation of this scalar solution would yield:
\begin{equation}
\vec{I} (z) = \int_{z_0}^{z} e^{-\int_t^z \tens{K} dt'} \vec{J}(t) dt +  
e^{-\int_{z_0}^z \tens{K} dt'} \vec{I}(z_0).
\label{fss}
\end{equation}
Note that, in this expression, all the mathematical operations involving 
matrices are well defined. For instance, the exponential of
a square matrix is defined as
\begin{equation}
 e^{\tens{L}} = \bbbone + \tens{L} +\frac{1}{2!}\tens{L}^2+ 
\frac{1}{3!}\tens{L}^3+\ldots 
\label{def}
\end{equation}
Now, just one difference arises between  algebras using matrices and scalars:
while two scalars always commute, that is not true in general for two matrices:
 $$\tens{K} \tens{L} \neq \tens{L} \tens{K}. $$
This difference becomes important when treating for example the derivative of
a power of a matrix: it is not true in general that
$$ \frac{d}{dz} \tens{L}^n = - n \tens{K} \tens{L}^{n-1}, $$
where we have used (and so we will do hereafter) that
$$ \frac{d}{dz}\tens{L} = - \tens{K}.$$
As a further consequence of the non-commutativity of matrices, we have that
$$  \frac{d}{dz}  e^{\tens{L}} \neq - \tens{K} e^{\tens{L}},$$
in complete contradiction with the scalar case.

In short, Eq.(3) is a  solution of the polarized RTE, Eq.(1), when the
commutation condition,
\begin{equation}
\left[ \tens{K},\tens{L} \right] = \tens{K} \tens{L} - \tens{L} \tens{K} =0,
\label{CC}
\end{equation}
holds. Under this assumption, the previous expressions recover  
an usual scalar appearance, and therefore we are permitted to write a 
scalar-like formal solution. It is interesting to note that
this condition does not imply a constant absorption matrix. In the following 
sections we 
will show how to incorporate variations  of  $\tens{K}$ with optical depth. 
Indeed, Landi 
Degl'Innocenti \& Landi degl'Innocenti \cite*{LDLD81,LDLD85} have already 
shown how to handle matrices of the form
$$ \tens{K}=\tens{K'}f(z)$$
with $\tens{K'}$ being a constant matrix. Although not explicitly said in 
these papers, it is 
obvious that  here $\tens{K}$ satisfies condition (\ref{CC}).

For the more general case when the commutation condition (\ref{CC}) does not
hold, Landi degl'Innocenti \& Landi degl'Innocenti 
\cite*{LDLD85} have derived a formal solution for the RTE:
\begin{equation}
\vec{I}(z)=\int_{z_0}^z \tens{O}(z,z')\vec{J}(z')dz' +
 \tens{O}(z,z_0)\vec{I}(z_0) ,
\label{fs}
\end{equation}
where  $\tens{O}(z,z')$ is the {\em evolution operator}, a new 
$4\times 4$ matrix which obeys the homogeneous equation
\begin{equation}
\frac{d}{dz}\tens{O}(z,z_0)= - \tens{K}(z)\tens{O}(z,z_0),
\label{eeq}
\end{equation}
with initial condition
$$ \tens{O}(z_0,z_0)=\bbbone ,$$ where $\bbbone$ represents the $4\times 4$ 
identity matrix. Note that when solution (\ref{fss}) applies, the evolution
operator takes an explicit form, namely:
\begin{equation}
\tens{O}(z,z')= \exp ({\tens{L}(z,z')})=
\exp \left(-{\int^z_{z'}\tens{K}dt}\right)
\label{cs}
\end{equation}
A general method to solve equations of the form of Eq. (\ref{eeq}) for linear
operators has already 
been  given by Magnus \cite*{Mag54}. In this remarkable paper an 
exponential expression is proposed:
$$\tens{O}(z,z') = \exp \left({\Omega (z-z')}\right), $$
were the exponent $\Omega(z-z')$ is given by an infinite series.

 Equation (\ref{cs}) turns out to be  Magnus´ expression when only the 
first term in the infinite series is kept (indeed the only non--zero one  
when condition (\ref{CC}) holds).

We now discuss three existing options to solve Eq.(\ref{RTE}):
\begin{enumerate}
\item  Constant matrix assumption. Condition (5) is inmediately satisfied and 
analytical solutions were found \cite{Unn56,Rac67}.
\item  Multi-layer techniques. The atmosphere is considered  as made up by
numerous successive  layers. 
A crude assumption on the radiative transfer in each optically thin layer is 
then advanced, and leads to a procedure of numerical integration expected by
intuition to converge to the exact solution. A formal proof of convergence 
was not given, but the numerical tests were indeed satisfactory. See for 
instance Rees \cite*{Ree87}, Rees {\em  et al.}\cite*{RM89}, 
Ruiz Cobo \& del Toro Iniesta \cite*{RCdTI92},
del  Toro Iniesta \& Ruiz Cobo \cite*{dTIRC96a}.
\item  Magnus´ solution. By applying linear algebra one can  treat the general 
case, with non commuting matrices \cite{Mag54}.
\end{enumerate}
The constant matrix technique, method (1), is not possible  when one 
wants to abandon the homogeneous magnetic field and atmosphere assumptions.
Next, poor economy is the main drawback of method (2). There is some doubt 
whether one can  determine {\em a priori} the number of layers necessary for 
a desired precision. Last but not least, a more analytical insight than what 
a pure numerical  method can give is always desired 
as well. Moreover, our ultimate purpose is magnetometry of the  sun or stars. 
We want to go beyond the first method, the most used, at present, but limited 
to a constant absorption matrix and therefore also constant field. Still we 
must admit that actual observations
 will allow us to  determine only ´little´ more than a homogeneous atmosphere 
model, say,   at most the magnetic fields at two or three levels in the 
atmosphere. It is therefore not ``economic''
to calculate more than a few layers in the atmosphere.

It is striking that Magnus' solution was published two years before the 
memorial paper by Unno\cite*{Unn56}, the first paper on RTE for polarized 
light, and as yet it has never been mentioned in the astrophysical 
literature. It is therefore given in Appendix A. The solution given by Magnus 
is mathematically exact,
but it requires the use of Lie algebra, is not economic and can hardly be 
used in practical 
computation. It is mentioned here because it confirms the approach of the 
present paper and complete it.

Our general strategy is, first, to  ``satisfy'' condition (5) as far as 
possible by extracting from the absorption matrix everything that commutes with
its integral and therefore can easily be integrated according to Eq.(3), 
as explained in section 2. In section 3, we diagonalise the commutative part 
of the matrix to allow an efficient integration. 
Then, in section 4, we treat the residual  matrix by an appropriate 
approximation  and thus obtain a semi-analytical
solution for an optically {\it finite} layer with arbitrary depth variations.  
Eventually we can then  borrow the techniques from the multi-layer approach 
and apply our semi-analytical solution to a few layer model to improve the 
computation.

A few words  on the mathematical space where we are working and where the RTE 
is to be solved, are in order. 
Magnetometry concerns the 3D  real physical space, where the magnetic field 
can be represented as a 3D ``vector''
 and all physical parameters of the atmosphere determine
the coefficients that enter the radiative transfer equation.
 The last one is much better calculated in another space. Indeed, we have 
already entered 
another  4D geometry: the Minkowski space, where the Stokes' 4--vectors are 
best described.
In this geometry, the norm of a 
vector $ \vec{I}$ is given by $(I^2-Q^2-U^2-V^2)^{\frac{1}{2}}$.
It has particular symmetries and is governed by linear algebra. The elementary 
operations, like absorption and retardation, are presented by matrices  for 
which commutation relations are of particular importance. 
When condition (5) holds, an exponential solution, scalar like, to a linear 
equation can easily be derived. Otherwise, we have to turn to 
Magnus' exponential solution.

The main difficulties originate from the fact that only few variables  are 
{\bf explicitly } common to the  ``two spaces''. Typically scalar variables 
like $z,k_c,K_l$ (see section 2. for their definitions) will appear 
 in both spaces in similar ways. However, rotations of the Stokes reference 
system will not.
Exception is the azimuth rotation. The  angle of rotation of the azimuth of 
the magnetic field in the ``real 3D 
space''  corresponds to a rotation in the Minkowski space, but with a double 
amount. Naturally, when a constant atmosphere is selected in the real space, 
the corresponding matrix in the Minkowski
space will be constant as well. On the other hand some rotation in the 
Minkowski space may be much easier to handle.
For instance, one may find convenient to use generalized Stokes vectors 
expressed in terms of elliptic states 
of polarization. Transformations from one set of Stokes reprensentation  to 
another are expressed simply as
 rotations in the Minkowski space. Except in some limiting cases it is not 
possible to translate these angles in terms of angles in the physical space. 
At the same time, the highly non linear relations between magnetic field and 
the entries of the absorption matrix cannot in general  be simplified. Thus, 
while the RTE can be solved for a given depth 
variation of the absorption matrix, we cannot, in general, recover 
analytically the corresponding variation of 
the magnetic field. We anticipate that numerical methods can overcome this 
difficulty and  profite from the 
analytical solution in the Minkowski space to treat the depth variations of 
the magnetic fields and improve both the 
economy and the precision of the calculations.   
These considerations apply as well  to all 
other atmospheric conditions, like temperature, pressure, velocity etc.    

In some particular cases, the relations between  variables in the Minkowski
and  real spaces may become simplified. For instance, in absence of absorption
of linear polarization, whether in the pure longitudinal magnetic field, or 
alternatively 
for particular Zeeman patterns, free of linear polarization. Also for the 
case when all Zeeman components
are separated, simple relations hold as will be discussed in the 
corresponding sections.

\section{Transformation of  matrix $ \tens{K} $ }

We rewrite the transfer equation as
$$\frac{d}{dz}\vec{I_0}= -\tens{K_0}\vec{I_0} + \vec{J_0}$$
And, $\tens{K_0}$ being invertible, we can define the 
source function vector, either LTE or not,
$$\vec{S_0}=\tens{K_0^{-1}}\vec{J_0},$$
so that the transfer equation reads
\begin{equation}
\frac{d}{dz}\vec{I_0}= -\tens{K_0}\left( \vec{I_0} - \vec{S_0} \right) .
\label{RT0}
\end{equation}
Matrix $\tens{K_0}$ can be decomposed as follows

\begin{eqnarray}
\tens{K_0}& = \kappa _l(z)  \pmatrix{
  0 & b \cos 2\phi & b \sin 2\phi & c        \cr
  b \cos 2\phi & 0 & \gamma_{\circ} & - \beta \sin 2\phi   \cr
  b \sin 2\phi  & - \gamma_{\circ} & 0 & \beta \cos 2\phi     \cr
  c    &   \beta \sin 2\phi   & - \beta \cos 2\phi  & 0} + \nonumber \\
 & +(g\kappa _l(z)+\kappa _c(z)) \bbbone
\label{K0}
\end{eqnarray}

 Where $ \kappa_l(z) $ and $\kappa_c(z)$ are the usual scalar absorption 
coefficients:
the selective (at line center) and the continuum one, respectively;
$ \phi $ is the azimuth angle of the magnetic field, relative to a
fixed reference system, and $\bbbone$ is the $4\times 4$ identity matrix.

This is the general symmetry of $ \tens{K_0} $; the
meaning of parameters $ g,b,c, \beta $,and $ \gamma_{\circ} $ can be 
found by com\-pa\-ring expression (\ref{K0}) with the corresponding ones in
Landi Degl'In\-no\-cen\-ti \& Landi Degl'Innocenti\cite*{LDLD81,LDLD85}, 
Rees \cite*{Ree87} or Ka\-wa\-ka\-mi\cite*{Kaw83}. 

We can simplify this matrix by rotating it an  angle $2\phi$ in the 
plane Q--U. That is, we introduce a rotation matrix
$$
\tens{R_1}= \pmatrix{1 & 0 & 0 & 0 \cr
0 & \cos 2\phi & \sin 2\phi & 0  \cr
0 & - \sin 2\phi & \cos 2\phi & 0 \cr
0 & 0 & 0 & 1} $$
and its inverse $\tens{R_1^{-1}}$, and we apply them to $\tens{K_0}$ to obtain
\begin{eqnarray}
 \tens{K_1'}=\tens{R_1}\tens{K_0}\tens{R_1^{-1}}= \kappa _l(z)    \pmatrix{
  0 & b & 0 & c        \cr
  b & 0 & \gamma_0 & 0 \cr
  0 & - \gamma_0 & 0 & \beta    \cr
  c    &   0  & - \beta   & 0} +\nonumber \\
 +(g\kappa _l(z)+\kappa _c(z)) \bbbone . 
\end{eqnarray}
Applying this transformation to Eq. (\ref{RT0}) we obtain:
\begin{equation}
\tens{R_1}\frac{d}{dz}\vec{I_0}=- \left(\tens{R_1}\tens{K_0}\tens{R_1^{-1}}
\right)\tens{R_1}\vec{I_0}+\tens{R_1}\vec{J_0}= -  \tens{K_1'} 
\vec{I_1}+\vec{J_1}
\label{E9}
\end{equation}
where
$$ \vec{I_1}=\tens{R_1}\vec {I_0},$$
$$ \vec{J_1}=\tens{R_1}\vec {J_0}.$$
The left hand side of the transformed transfer equation (\ref{E9}) is equal to
$$ 
\frac{d}{dz}(\tens{R_1}\vec{I}_{\circ})-(\frac{d}{dz}\tens{R_1})\vec{I}_{\circ}
.$$
where we note that
\begin{eqnarray}
(\frac{d}{dz}\tens{R_1})\vec{I_0}=(\frac{d}{dz}\tens{R_1})\tens{R_1^{-1}}
\tens{R_1}\vec{I_0}= \nonumber \\
=2\left(\frac{d\phi}{dz}\right)\pmatrix{
  0 &  0 & 0 & 0    \cr
  0 &  0 & 1 & 0    \cr
  0 & -1 & 0 & 0    \cr
  0 &  0 & 0 & 0} \vec{I_1},
\end{eqnarray}
so that we can write the transformed transfer equation  as
\begin{equation}
\frac{d}{dz}\vec{I_1}=-\tens{K_1}\vec{I_1}+\vec{J_1}
\label{RT1}
\end{equation}
where 
$$\tens{K_1} =\kappa _l(z) \tens{N_1} +(g\kappa _l(z)+\kappa _c(z))\bbbone ,$$ 
with
$$ \tens{N_1}=\pmatrix {0 & b & 0 & c        \cr
  b & 0 & \gamma & 0 \cr
  0 & - \gamma & 0 & \beta    \cr
  c    &   0  & - \beta   & 0},
$$
where we have introduced
\begin{equation}
\gamma = \gamma_0-2\frac{d\phi}{dz}\frac{1}{\kappa _l} .
\label{gamma}
\end{equation}
The meaning of the new Stokes reference system is as follows: after the 
$ \phi$ rotation, the new generalized Stokes parameters $ Q_1 $ and $ U_1$,  
projections
of vector $\vec{I}$  on axis $\bf{Q_1}$ and $\bf{U_1}$ in the new reference 
system, still correspond to linear polarization, but Zeeman linear absorption 
affects $ Q_1$ only (absorption along the $ \bf{Q_1}$ axis). Faraday rotation 
may still affect $U_1$, but with zero absorption. The parameters 
 $I_1$ and $V_1$ are unchanged (the correspoding axis $\bf{I}$ and $\bf{V}$ 
are not affected by the $\phi$ rotation). In the real space, the meaning of 
this rotation is that the reference for
the usual definition of the Stokes parameters is taken parallel  to the 
magnetic field for $Q$. These new axes rotate with the field.

A second simplification is obtained by the use of a new rotation, given by 
\begin{equation}
\tens{R_2} = \pmatrix{
  1 &  0   & 0  & 0        \cr
  0 &  \cos \alpha   & 0  & \sin \alpha        \cr
  0 &  0   & 1  & 0        \cr
  0 & -\sin \alpha   & 0  & \cos \alpha}
\end{equation}
and its inverse $\tens{R_2}^{-1}$, where
\begin{eqnarray}
\cos \alpha = \frac{b}{\sqrt{b^2 + c^2}} \nonumber \\
\sin  \alpha = \frac{c}{\sqrt{b^2 +  c^2}}.
\end{eqnarray}
By applying it to the matrix $\tens{K_1}$ we obtain
\begin{eqnarray}
\tens{K_2'}= \tens{R_2}\tens{K_1}\tens{R_2^{-1}}= \kappa _l(z)    \pmatrix{
  0 & q & 0 & 0        \cr
  q & 0 & p' & 0 \cr
  0 & -p' & 0 & r'    \cr
  0    &   0  & - r'   & 0} + \nonumber \\
 +(g\kappa _l(z)+\kappa _c(z)) \bbbone ,
\end{eqnarray}
where \begin{eqnarray}
q &=& \sqrt{b^2 + c^2} ,\nonumber \\
p' &=& \frac{\gamma b - \beta c}{q} ,\nonumber \\
r' &=& \frac{\gamma c + \beta b}{q}.
\end{eqnarray}
For the singular case $ q=0 $, we adopte $\alpha=0$, $p'=\gamma$ and 
$ r'=\beta$.

The meaning of the new Stokes reference system is as follows: after the 
$ \alpha$ rotation, the new generalized Stokes parameters $ Q_2 $ and $ V_2$, 
projections of $\vec{I_1}$ on axis $\bf{Q_2}$ and $\bf{V_2}$ 
correspond to elliptic polarizations, Zeeman elliptic absorption affects $Q_2$ 
only (absorption along axis $ \bf{Q_2}$). Faraday rotation still affects $V_2$
(and $U_2$) but with zero absorption. Parameters  $I_2$ and $U_2$ are 
unchanged (the correspoding axis
$\bf{I}$ and $\bf{U}$ are not affected by the $\alpha$ rotation). Note that 
$\alpha$ is wavelength
dependent and  therefore the rotation in the Minkowski space is not constant 
with  $ \lambda$!

In the real space, the meaning of this rotation is not any longer as simple as 
before. However, note that
the most general state of polarisation is elliptic!! At each $\lambda $ we 
can determine 
the ellipse of polarisation absorbed by the Zeeman effect. We then choose it 
as axis ${\bf Q_2}$. 
The complete new generalized Stokes system follows from transformation matrix 
$\tens{R_2}$.
In deriving $ \alpha $, $ q=\sqrt{b^2 + c^2} $ stands for the total intensity 
of the elipse of polarisation and 
$ \sin  \alpha = \frac{c}{\sqrt{b^2 +  c^2}}$  is the rate 
of circular polarisation. Although easy to calculate, $\alpha$ has no simple 
meaning in terms of the magnetic field, except for the case of a strong field 
when all the Zeeman components are completely separated.
Then $\alpha=0$ for the $\pi$ component, and $\tan\alpha= \pm \frac{2 \cos 
\theta}{\sin^2\theta} $ for the $\sigma$ components, where $ \theta$ is the 
inclination angle of the magnetic field.

We repeat here all the steps made for first rotation $\tens{R_1}$, defining 
the new transformed Stokes and emission vectors
\begin{eqnarray}
\vec{I_2}=\tens{R_2}\vec{I_1} \nonumber \\
\vec{J_2}=\tens{R_2}\vec{J_1} ,
\end{eqnarray}
and calculating the term
\begin{equation}
\left(\frac{d}{dz}\tens{R_2}\right) \tens{R_2^{-1}}= \left(\frac{d\alpha}{dz}
\right)
\pmatrix{0&0&0&0\cr 0&0&0&1 \cr 0&0&0&0\cr 0&-1&0&0}
\end{equation}
 so that we can write the transformed transfer equation as
\begin{equation}
\frac{d}{dz}\vec{I_2}=-\tens{K_2} \vec{I_2} +\vec{J_2},
\label{RT2}
\end{equation}
where 
$$ \tens{K_2}=\kappa _l(z)\tens{N_2}+(g\kappa _l(z)+\kappa _c(z))\bbbone ,$$ 
with
$$ \tens{N_2}=\pmatrix {0 & q & 0 & 0        \cr
  q & 0 & p' & -s \cr
  0 & - p' & 0 & r'    \cr
  0    &   s  & - r'   & 0},
$$ 
where we have introduced a new parameter
$$ s = + \frac {d\alpha }{dz}\frac{1}{\kappa _l} .$$
This parameter $s$, a new non-zero entry in $\tens{K_2}$, 
makes it a little more 
complicate than before. A new transformation is necessary if 
we want to obtain a simpler matrix like $\tens{K_2'}$. The way for this
simplification is a third rotation $\tens{R_3}$ given by 
\begin{equation}
\tens{R_3} = \pmatrix{
  1 &  0   & 0  & 0        \cr
  0 &  1  & 0  & 0        \cr
  0 &  0   & \cos \xi  & -\sin \xi        \cr
  0 & 0   & \sin \xi  & \cos \xi},
\end{equation}
with
\begin{equation}
\cos \xi = \frac{p'}{\sqrt{p'^2+s^2}} \\  \sin \xi =\frac{s}{\sqrt{p'^2+s^2}} .
\end{equation}
The same mathematical steps of the two previous rotations are repeated for 
$\tens{R_3}$. We pass directly
to the final expression for the transfer equation:
\begin{equation}
\frac{d}{dz}\vec{I_3}=-\tens{K_3}\vec{I_3} + \vec{J_3}        
\end{equation}
where $\tens{K_3}$ has the following aspect:
$$\tens{K_3} =\kappa _l(z) \tens{N_3} +(g\kappa _l(z)+\kappa _c(z)) \bbbone ,$$
with
$$ \tens{N_3}=\pmatrix {0 & q & 0 & 0        \cr
  q & 0 & p & 0 \cr
  0 & - p & 0 & r    \cr
  0    &   0  & - r   & 0}
$$ 
where we have defined
\begin{eqnarray}
p =\sqrt{p'^2+s^2} \nonumber \\
r =r'-\frac{d\xi}{d z}\frac{1}{\kappa _l}.
\end{eqnarray}

By now the matrix, and consequently the RTE, has been simplified in a general
way, without any assumption nor constraint. To proceed to an exponential  
solution for 
the  transfer  equation, we need to ensure that  commutation condition 
(\ref{CC}) holds for $\tens{K_3}$. A necessary and sufficient
condition for that is a matrix $ \tens{N_3}$ of the form
\begin{equation}
\tens{N_3}= (\tens{N_3})_0 \cdot f(z)
\label{N3}
\end{equation}
where $(\tens{N_3})_0$ is a constant matrix, and $f(z)$ is any scalar function
of $z$. Let $P$,$Q$ and $R$ be the integrals of $p$,$q$ and $r$. When 
calculating the commutator of $ \tens{N_3}$ with its integral 
$\left[\tens{N_3},\tens{L_3}\right]$, its only 
{\it a priori} non--zero entries are  
$(Pq-Qp)$ or $(Pr-Rp)$. It is very easy to see that 
these expressions vanish when 1) $p=0$, 2) $q=r=0$ or 3) $p$,$q$
 and $r$ are all proportional to the same scalar function $f(z)$. 
While the treatement of cases
1) and 2) is straightforward, the general case 3) needs some discussion:  we
rewrite $p$,$q$  $r$ as $p_0 f(z)$, $q_0 f(z)$ and $r_0 f(z)$.
Matrix $(\tens{N_3})_0$ keeps the appearance of 
$ \tens{N_3}$ but with $p$, $q$ and $r$ substituted by $p_0$,$q_0$ and
$r_0$. These new variables to be constant is the
{\bf necessary and sufficient condition} for writing
an exponential scalar-like solution. A constant  $\tens{K}$ matrix implies 
7 constant parameters.
Matrix $\tens{N_3}$ contains only three parameters $q,p$ and $r$, but we need 
to keep constant only the two ratios $p/q$ and $r/q$ to satisfy condition (5).
After transformation, only two variables are requested to be
constant,i.e., two constraints instead of seven originally: 5 degrees of 
freedom  have been earned. Reviewing the transformation
process, those 5 degrees of freedom may be used to treat analytically 
gradients with depth in azimuth, $\kappa _c$ and $\kappa _l$, angles
$\alpha$ and $\xi $, and $f(z)$. Details about how to do it 
 and its application to a numerical code, are left for a forthcoming paper.

 For the sake of demonstration, we discuss the variation of angles  $\alpha$ 
and  $\theta$ alone in  the case of separated Zeeman components. 
In absence of azimuth variation $p'=0$ and $p=s$.  

For the $\pi$ Zeeman  component, polarisation is purely linear, 
 $\alpha=0$, and we adopt $ \xi=0$ $p=s=0$: the integration is 
straightforward (case 1).
For the $\sigma$ Zeeman components, polarisation is elliptic. Since $p'=0$, we 
adopt $ \xi=\pi/4$ and $r=r'$.
$$  \frac {d\alpha }{dz}= \pm\frac{1}{4}\frac{d(\cos\theta)}{dz}\frac{2}
{1+\cos^2\theta}.$$
Absorption in each $\sigma $ component is proportional to 
$q \propto 1+ \cos^2\theta$
and also $r=r' \propto 1+ \cos^2\theta $, $\theta$ being the only variable 
depending on depth.
We now suggest $f(z)=1+\cos^2\theta$ to match the depth variation of $q$, $r$
and $p=s= \frac {d\alpha }{dz}\frac{1}{\kappa _l}$ (case 3) with
$ \xi=\pi/4$
To keep the same depth variation for $\frac {d\alpha }{dz}$, we impose a 
variation of $\theta$ such that
$$\frac{d(\cos \theta)}{dz}\frac{1}{(1+\cos^2\theta)^2} =\mbox{constant},$$
and we obtain $$ \frac{\cos\theta}{1+\cos^2\theta}+\arctan(\cos\theta)=
2 \: \mbox{constant}(z-z_0)$$                      
In general, it will be impossible to interpret $\alpha$ in such a simple way.

\section{Diagonalization of the off-diagonal matrix}

In the last section, matrix $\tens{K}$ was simplified as much as possible. All
the way to the final form we have seen how to incorporate some atmospheric
gradients into the exponential solution, and  at the end we have required
the commutation condition to still hold with minimum freedom restrictions. 
We can then calculate the 
matrix in the exponent of the solution; but we must compute the 
exponential of this matrix too. The last can be done by using the matrix series
(\ref{def}), but we prefer the comfort of calculating the 
exponential of scalars. We can achieve this purpose by diagonalizing 
$\tens{K_3}$. In fact, we only need to diagonalize
$(\tens{N_3})_0$. We solve  the eigenvalue equations for 
this matrix:
$$(\tens{N_3})_0\vec{U}_i=\lambda_i \vec{U}_i$$
The four eigenvalues $ \lambda_i $ are given by:
\begin{equation}
\lambda _i^4+\lambda _i^2(r_0 ^2+p_0 ^2-q_0 ^2)-q_0 ^2r_0 ^2=0
\end{equation}
which is a biquadratic equation, so first we solve
\begin{equation}
\lambda _i^2= \frac{(q_0 ^2-r_0 ^2-p_0 ^2) \pm \Delta}{2}
\end{equation}
where
\begin{equation}
\Delta ^2= (r_0 ^2+p_0 ^2-q_0 ^2)^2+4q_0 ^2r_0 ^2 ,
\end{equation}
and the four eigenvalues are therefore:
\begin{eqnarray} 
\lambda _1 &=&-\sqrt{(q_0 ^2-r_0 ^2-p_0 ^2+\Delta )/2}, \\
\lambda _2&=& - \lambda _1 ,\\
\lambda _3&=&- \sqrt{(q_0 ^2-r_0 ^2-p_0 ^2-\Delta)/2} ,\\
\lambda _4&=&- \lambda _3 .
\end{eqnarray}
We introduce the notations:
\begin{eqnarray}
\theta_i&=& -\lambda _i    \\
\delta_i&=&\lambda _i^2-q_0 ^2 ,
\end{eqnarray}
noting that
\begin{eqnarray*}
\theta_1= -\theta_2  \nonumber \\
\theta_3= -\theta_4 ,
\end{eqnarray*}
and that
\begin{eqnarray*}
\delta_1=\delta_2    \nonumber \\
\delta_3=\delta_4 .
\end{eqnarray*}
Using this notation, one finds that the corresponding eigenvectors  
$ \vec{U}_i $ are given by
\begin{equation}
\vec{U}_i=\frac{1}{\sqrt{2} q_0 } \pmatrix{q_0    \cr -\theta_i  \cr 
\delta _i/p_0     \cr  r_0 \delta _i/(p_0 \theta _i) } .
\end{equation}
We may now calculate matrices $\tens{T}$ and $ \tens{T^{-1}} $  which 
diagonalize  $(\tens{N_3})_0$:
\begin{equation}
\tens{T^{-1}}=\frac{1}{q_0 p_0  \sqrt{2} } \pmatrix{
q_0 p_0  & q_0 p_0  & q_0 p_0  & q_0 p_0     \cr
-p_0 \theta_1  & p_0 \theta_1  & -p_0 \theta_3  & p_0 \theta_3    \cr
\delta_1    & \delta_1    & \delta_3    & \delta_3    \cr
r_0 \delta_1/\theta_1 & -r_0 \delta_1/\theta_1 &
r_0 \delta_3/\theta_3 & -r_0 \delta_3/\theta_3}
\end{equation}
and 
\begin{equation}
\tens{T}= \frac{1}{\Delta \sqrt{2}} \pmatrix{
-\delta_3  &   \theta_1 \delta_3/q_0    &  q_0 p_0   &
 \theta_1 \theta_3^2p_0 /(r_0 q_0 )     \cr
-\delta_3  &  -\theta_1 \delta_3/q_0    &  q_0 p_0   &
- \theta_1 \theta_3^2p_0 /(r_0 q_0 )     \cr
\delta_1  &   -\theta_3 \delta_1/q_0    &  -q_0 p_0   &
- \theta_3 \theta_1^2p_0 /(r_0 q_0 )     \cr
\delta_1  &   \theta_3 \delta_1/q_0    &  -q_0 p_0   &
 \theta_3 \theta_1^2p_0 /(r_0 q_0 )     }.
\end{equation}
These transformation matrices $\tens{T}$ and $\tens{T^{-1}}$ just derived can 
be
 applied in general, save for those exceptions where they become singular. 
These particular cases are:\begin{enumerate}
\item The case when $p_0 =0$.

To solve this singularity we suggest the following transformation matrix and 
its inverse as a substitution of the previous ones:
\begin{equation}
\tens{T}= \frac{1}{\sqrt{2}}\pmatrix{
1 & -1 & 0 & 0    \cr
1 & 1 & 0 & 0    \cr
0 & 0 & 1 & i    \cr
0 & 0 & i & 1} ,
\end{equation}
with the resultant diagonalization of $ (\tens{N_3})_0 $ :
\begin{equation}
\tens{T}(\tens{N_3})_0\tens{T}^{-1}=  \pmatrix{
-q_0  & 0 & 0 & 0    \cr
0 & +q_0  & 0 & 0    \cr
0 & 0 & -ir_0  & 0    \cr
0 & 0 & 0 & +ir_0 } .
\end{equation}

\item The case when $q_0 =0$.

With no loss of generality one can take $ \alpha=0 $ and then $\tens{R_2}$ (and
subsequently $\tens{R_3}$) becomes the identity matrix. For the 
diagonalization transformation,   the eigenvalues are the following:
\begin{eqnarray} 
\lambda_1 &=& 0 ,\\
\lambda_2 &=& 0 ,\\
\lambda_3 &=&  - i \Delta^{\frac{1}{2}} ,\\
\lambda_4 &=&  i \Delta^{\frac{1}{2}}.
\end{eqnarray}
Here $ \Delta $ is reduced to
\begin{equation}
\Delta = p_0 ^2+ r_0 ^2 ,
\end{equation}
and the diagonalizing matrix is :
\begin{equation}
\tens{T}=\frac{1}{2\Delta} \pmatrix{
2\Delta & 0 & 0 & 0    \cr
0 & 2r_0  & 0 & 2p_0     \cr
0 & p_0  & i\Delta^{\frac{1}{2}} & r_0     \cr
0 & p_0  & -i\Delta^{\frac{1}{2}} & -r_0 }.
\end{equation}

\item The case when $r_0 =0$, and $ q_0 ^2\neq p_0 ^2 $. 

 In this case, $ \Delta =q_0 ^2-p_0 ^2 $ ; $ \Delta^{\frac{1}{2}} $ is 
imaginary if $ q_0 ^2<p_0 ^2 $.
The eigenvalues are given as follows:
\begin{eqnarray} 
\lambda_1 &=&  i \Delta^{\frac{1}{2}} ,\\
\lambda_2 &=& - i \Delta^{\frac{1}{2}} ,\\
\lambda_3 &=& 0 ,\\
\lambda_4 &=& 0 ,
\end{eqnarray}
and the transformation matrix 
\begin{equation}
\tens{T}=\frac{1}{2\Delta}\pmatrix{
q_0  & i\Delta^{\frac{1}{2}} & p_0  & 0    \cr
q_0  & -i\Delta^{\frac{1}{2}} & p_0  & 0    \cr
-2p_0  & 0 & -2q_0  & 0    \cr
0 & 0 & 0 & 2\Delta}.
\end{equation}

\item The case when $ r_0 =0 $, and $ p_0 =q_0  $. 

This is the only case where matrix $(\tens{N_3})_0$ cannot be diagonalized. 
It will be reduced by using
\begin{equation}
\tens{T}= \pmatrix{
1 & 0 & 0 & 0    \cr
0 & 1 & 0 & 0    \cr
1 & 0 & 1 & 0    \cr
0 & 0 & 0 & 1}
\end{equation}
and its inverse, so that we get
\begin{equation}
\tens{T}(\tens{N_3})_0\tens{T}^{-1}=
\pmatrix{
0 & q_0  & 0 & 0    \cr
0 & 0 & q_0  & 0    \cr
0 & 0 & 0 & 0    \cr
0 & 0 & 0 & 0}.
\end{equation}

Note that the resulting matrix, although not diagonal is simple enough as
to have its exponential calculated as
\begin{equation}
\mbox{e}^{-\int \tens{T}\tens{N_3}\tens{T}^{-1}dt}= \pmatrix{
1 & -Q & Q^2/2 & 0    \cr
0 & 1 & -Q & 0    \cr
0 & 0 & 1 & 0    \cr
0 & 0 & 0 & 1},
\end{equation}
where we have used the notation $ Q =\int q dt $.
\end{enumerate}

Back to the general case,  we diagonalize (or apply the equivalent 
transformation in case 4) to  the RTE . It is  a new transformation of the 
equation, as were $\tens{R_1}$, $\tens{R_2}$ and $\tens{R_3}$, and as in these 
3 rotations, a term of the form 
$$ \left( \frac{d}{dz}\tens{T}\right) \tens{T^{-1}}\tens{T} \vec{I_3}$$
appears in the transformed equation. However, as all the entries of 
$(\tens{N_3})_0$, and subsequently of $\tens{T}$,  are constant, this term is 
immediately zero, in perfect 
agreement with the statement that expression (\ref{N3}) is a necessary and 
sufficient  condition to write a scalar--like exponential 
solution.\footnote{ If we 
relax condition  (\ref{N3}), $\tens{N_3}$ can still be diagonalized by 
substituting $p,q$ and $r$ for $p_0 , q_0 $ and $r_0$ in $\tens{T}$ and
 $\tens{T}^{-1}$. But this time
  $\frac{d}{dz}\tens{T}$ is not zero. We cannot get rid of this term by 
mathematical manipulations, the RTE cannot be diagonalized and a scalar-like
solution as Eq. (\ref{fss}) cannot be applied.}

For the general case, we define the new generalized 
Stokes and emission vectors	after the diagonalizing transformation
for the RTE  as
\begin{eqnarray}
\vec{I}_T=\tens{T} \vec{I_3} \nonumber \\
\vec{J}_T=\tens{T} \vec{J_3}
\end{eqnarray}
and rewrite the diagonalized transfer equation as
\begin{equation}
\frac{d}{dz} \vec{I}_T = - \tens{\Lambda}\vec{I}_T+\vec{J}_T        
\label{DTE}
\end{equation}
where 
$$ \tens{\Lambda}= \kappa _l (z) \tens{T} (\tens{N_3})_0 \tens{T^{-1}}f(z) 
+ (g\kappa_l (z)+ \kappa _c (z))\bbbone$$
is the diagonalized matrix. 
For the last singular case, the equivalent reduced matrix should be used.

\section{The formal solution in a particular atmosphere}

In the previous two sections, we have seen how to transform matrix 
$\tens{K}$ in order to be able to write an exponential solution in a 
convenient form, and to be able
to calculate that exponential as easily as the scalar case. We have shown 
that this procedure  
applies to absorption matrices whose depth dependence has a few degrees of 
freedom but still satisfying condition (5). Summing up the last two sections,
 we can
manage gradients  in the azimuth $\phi$ of the magnetic field and in the angles
 $\alpha$ and $\xi$, variations in $\kappa _l$ and $\kappa _c$, treat 
analytically a general function $f(z)$ in the final matrix $\tens{N_3}$, and
integrate for any emission vector $\vec{J}$. 
 
In this context we assume an ideal atmosphere satisfying condition
(\ref{N3}). We write a formal solution (\ref{fss}) for the diagonalized 
transfer equation (\ref{DTE}): 
\begin{equation}
\vec{I}_T=\int_{z_0}^z \mbox{e}^{-\int_t^z \tens{\Lambda}(t')dt'}
\vec{J}_T(t)dt +
 \mbox{e}^{-\int_{z_0}^z \tens{\Lambda}(t')dt'}\vec{I}_T(z_0)
\label{solT}
\end{equation}
Except for the last singular case, we can apply a diagonal matrix  
$\tens{\Lambda}$, and write its elements as
\begin{equation} 
\tens{\Lambda}_{ii}=\kappa _l \lambda _i f(z)+ (g\kappa _l + \kappa _c)
\end{equation}
where the $\lambda _i$'s coincide with the eigenvalues of the off-diagonal
matrix $(\tens{N_3})_0$. We can now write a scalar disentangled solution for 
each
one of the four components of the generalized Stokes vector as
\begin{eqnarray}
(\vec{I}_T)_i (z)= \int_{z_0}^z \mbox{e}^{-\int_t^{z} (\kappa _l \lambda _i 
f(t')+ g\kappa _l + \kappa _c)dt'} (\vec{J}_T)_i (t)dt
+\nonumber \\
+ \mbox{e}^{-\int_{z_0}^{z}(\kappa _l \lambda _i f(t')+ g\kappa _l + \kappa _c)
dt'}(\vec{I}_T)_i (z_0) 
\end{eqnarray}
where we can define four generalized optical depths
$$
\tau _i (z,z_0)=\int_{z_0}^{z}(\kappa _l\lambda _i f(t')+ g\kappa _l + 
\kappa _c)dt' ,$$
complex in general,
with which the solutions are
\begin{equation}
(\vec{I}_T)_i (z)= \int_{z_0}^z \mbox{e}^{ \tau _i (z,t)}(\vec{J}_T)_i(t) dt+
\mbox{e}^{\tau _i (z,z_0)}(\vec{I}_T)_i (z_0) .
\end{equation} 
We can relate now the generalized Stokes vector $\vec{I}_T$ with the physical
Stokes vector $\vec{I_0}$ by writting in order all the transformations
that have been made, i.e.
\begin{equation}
\vec{I}_T=\tens{T}\tens{R_3}\tens{R_2}\tens{R_1}\vec{I_0}=\tens{T_T}\vec{I_0}
\end{equation}
where $\tens{T_T}$ is the complete transformation, product of all  others
in the proper  order. Evidently $\tens{T_T}$ is invertible, so that we can make
 back way from the generalized Stokes vector to the physical one:
$$\vec{I_0}=\tens{T_T^{-1}}\vec{I}_T .$$

\section{An approximation for a general atmosphere}
\label{num}

Model atmospheres are usually not so perfect as to be considered in the
cases treated in the last section.  In 
the absence of a general analytical solution for the evolution operator, we
must manage those general models numerically. Our strategy is to integrate
everything we can and to linearize the rest. For this purpose we borrow a 
technique from other successful numerical  integrators as the well-known DELO 
\cite{RM89}. In what follows we develop this idea.

We want to integrate the transfer equation
\begin{equation}
\frac{d}{dz}\vec{I}=-\tens{K}\vec{I} + \vec{J}
\end{equation}
or alternatively
\begin{equation}
\frac{d}{dz}\vec{I}=-\tens{K}\left( \vec{I} - \vec{S} \right).
\label{e1} 
\end{equation}
The integration is to be made
in the interval $z_1<z<z_2$ and the atmosphere in the two extreme 
points $z_1$ and $z_2$ is specified, that is, we know $\tens{K}(z_1)$ and 
$\tens{K}(z_2)$
 and also $\vec{S}(z_1)$ and $\vec{S}(z_2)$. The incoming light $\vec{I}(z_1)$
is also given. We want to obtain
the polarized light at $z_2$: $\vec{I}(z_2)$. After the last section we already
know how to integrate the transfer equation if the atmosphere is characterized
by the 
prescriptions given there. Note that, given the atmosphere at the two points
 $z_1$ and $z_2$, one can always calculate the angles $\alpha$ and $\xi$ and 
the matrix $\tens{N_3}$ at those two levels, and look for  a suitable 
integrable 
atmosphere in agreement with expression (\ref{N3}) which satisfy the data
at  $z_1$ and $z_2$ as far as possible. So let us approximate our atmosphere 
between these two levels by this integrable atmosphere, 
represented by a matrix $\overline{\tens{K}}$ and an emission vector 
$\overline{\vec{S}}$. We can obtain a solution $\overline{\vec{I}}(z_2)$ for 
the equation
\begin{equation}
\frac{d}{dz}\overline{\vec{I}}=-\overline{\tens{K}}(\overline{\vec{I}}-\overline{\vec{S}})
\label{e2}
\end{equation}
taking as initial condition $\overline{\vec{I}}(z_1)=\vec{I}(z_1)$.
Substraction of Eq. (\ref{e2}) from Eq. (\ref{e1}) results in
\begin{equation}
\frac{d}{dz}(\vec{I}-\overline{\vec{I}})= 
\tens{K}(\vec{I}-\vec{S})-\overline{\tens{K}}(\overline{\vec{I}}-
\overline{\vec{S}}),
\label{A3}
\end{equation}
and upon formal integration :
\begin{equation}
\vec{I}_2-\overline{\vec{I}}_2= 
\int^{z_2}_{z_1}\left(\tens{K}(\vec{I}-\vec{S})-
\overline{\tens{K}}(\overline{\vec{I}}-\overline{\vec{S}})\right)dt,
\label{A4}
\end{equation}
where $\vec{I}_2=\vec{I}(z_2)$ and $\overline{\vec{I}}_2=\overline{\vec{I}}
(z_2)$ given as solution to Eq. (\ref{e2}). This equation reflects the error
made under the previous approximation. Following our strategy, once we have 
solved for the integrable part we linearize
the rest. So that we now assume that the right hand side of 
equation (\ref{A3}) is 
small and can be  linearized in the interval $ z_1 < z < z_2$.
We define 
\begin{equation}
y=
\tens{K}(\vec{I}-\vec{S})-\overline{\tens{K}}(\overline{\vec{I}}-
\overline{\vec{S}})
\end{equation}
and upon linearization we write 
\begin{equation}
y=a+b(z-z_1),
\end{equation}
where 
\begin{equation}
a=\tens{K}_1(\vec{I}_1-\vec{S}_1)-\overline{\tens{K}}_1
(\overline{\vec{I}}_1-\overline{\vec{S}}_1).
\end{equation}
At $ z=z_2 $, we have $ y=a+b(z_2-z_1) $,  and we obtain
\begin{equation}
a + b(z_2-z_1)=\tens{K}_2(\vec{I}_2-\vec{S}_2)-\overline{\tens{K}}_2
(\overline{\vec{I}}_2-\overline{\vec{S}}_2).
\end{equation}
To solve Eq. (\ref{A4}) we write 
\begin{equation}
\vec{I}_2-\overline{\vec{I}}_2= \int^{z_2}_{z_1}ydt.
\end{equation}
And by means of the linearization the last integral becomes 
$$ a(z_2-z_1) + \frac{b}{2}(z_2-z_1)^2$$
so that, substituting $a$ and $b$ by its complete expressions
\begin{eqnarray}
\vec{I}_2-\overline{\vec{I}}_2 & = & 
\frac{z_2-z_1}{2} \left(\tens{K}_2(\vec{I}_2 - \vec{S}_2) - 
\overline{\tens{K}}_2(\overline{\vec{I}}_2 - \overline{\vec{S}}_2) \right. 
 \nonumber \\
& &\left. +(\tens{K}_1 - \overline{\tens{K}}_1)\vec{I}_1 - \tens{K}_1\vec{S}_1 
+\overline{\tens{K}}_1\overline{\vec{S}}_1\right) .
\label{A10}
\end{eqnarray} 
In this expression everything is already known except for $\vec{I}_2$ that is
precisely what we want to calculate.

A convenient choice of the overlined parameters may render equation (\ref{A10})
simpler. 
For illustration, let us  choose  
$$ \overline{\tens{K}}_1 = \overline{\tens{K}}_2 = \tens{K}_2 $$
and $ \overline{\vec{S}}_1 = \vec{S}_1 $, 
$ \overline{\vec{S}}_2 = \vec{S}_2 $.
We then obtain 
\begin{equation}
\vec{I}_2 =\overline{\vec{I}}_2 - \left[1-\frac{z_2-z_1}{2}
\tens{K}_2\right]^{-1}
\frac{z_2-z_1}{2} \left(\tens{K}_2 - \tens{K}_1\right) (\vec{I}_1 - \vec{S}_1),
\end{equation}
a solution for $\vec{I}(z_2)$. This solution is not exact,
its precision depends on how good the linear approximation is. In the limit,
we can made the integration interval $(z_1,z_2)$ as small as we want 
but at the cost of increasing the number of layers.
A compromise will be necessary between speed and required precision.

\section{Conclusions}

The purpose of this paper was first to deepen our understanding of the 
integration of the RTE for polarised light and next to improve the basis  
for numerical codes. The main conclusions is: {\em The fundamental key to 
solve the RTE of polarised light is the commutation of the absorption matrix 
and its integral.}

When this commutation condition  is satisfied : 
\begin{enumerate}
\item A scalar like solution can be proposed to the vector equation.
\item A constant absorption matrix satisfies the commutation requirement, 
 however, it is only a sufficient condition,  not necessary. 
After some elaboration one can show that only two constraints at all are 
necessary instead of the seven inherent in the fully constant absorption 
matrix.
\item In general, it will possible to diagonalize the absorption matrix
and consequently also the RTE with its vector solution. This results in four
scalar equations with four scalar solutions. The variables are 
no more the usual Stokes parameters, but generalized ones,  corresponding to 
general states of polarisation.
\item  The solution being analytical, it is valid for quite  thick optical
layers.  The real numerical application is beyond the scope of this paper.
 \end{enumerate}

When the commutation condition does not hold, one can turn to Magnus' solution,
described shortly in the appendix. 
Direct application of Magnus'  solution to a numerical code seems immature
at present.
For a general atmosphere, the numerical strategy proposed is to integrate
 analytically what we can and approximate the rest, that is:

\begin{enumerate} 

    \item Divide the atmosphere into a reasonably number of layers, so that
           in each of them the commutation condition is only slightly violated.
 \item    Approximate the general absorption matrix in each layer by an 
           average  that satisfies the commutation condition. 
\item     Apply the solution developed in this paper using last matrix.
 \item     Applying an approximation for the residual matrix,
           eventually the one used in DELO \cite{RM89}.
 \end{enumerate}

As an objective, we intend to improve the efficiency of integration and 
inversion codes. This will be a must in treating the abundant data expected
from  multi--line spectropolarimetric observations to be provided by
the French---Italian telescope THEMIS.

\appendix
\section{The general solution to the evolution operator}

We intend in this appendix to give a self-contained proof of 
Magnus' exponential solution
to the equation for the evolution operator. The original proof is
to be found in Magnus' paper\cite*{Mag54}. Here we will give  a short but
complete proof introducing 
in the meanwhile the necessary tools used while
working with Magnus' expressions and to get acquainted with algebraic
manipulations inherent to problems involving non-commutative operators.
Magnus makes use of an original technique to manage the derivative of the
exponential of a linear operator (a square matrix in our case). He transforms
this derivative into an algebraical expression, and uses it to solve the 
differential equation.

In what follows $w$, $x$, $y$ and $z$ stand for such operators also called
{\em Lie elements}. The {\em Lie product} of two Lie
elements $x$ and $y$ is defined by
$$w = \left[ x , y\right] ,$$
 where $w$ is a new Lie element. This product is usually called 
{\em commutator}. Now an abreviated notation is used for the l-fold 
Lie product by $x$ of $y$ as
$$ \{ y , x^l\}=[[\ldots [ y ,\overbrace{x]\ldots ,
x]}^{l \mbox{ times}}$$
with $\{ y , x^0\}=y$.
With this notation, it is easy to show by a straightforward calculation
that
\begin{equation}
\mbox{e}^{-x} w \mbox{e}^{x} = \sum _{l=0}^{\infty} \frac{1}{l!}\{w,x^l\},
\label{e-xyex}
\end{equation}  
where we remind that the exponential of an operator $x$ is defined as
$$\mbox{e}^{x} = \sum _{l=0}^{\infty} \frac{1}{l!} x^l .$$
Following Magnus, we extend the previous notation to a polynom $P(x)$ in an
evident form:
\begin{equation}
\{y,P(x)\}=\sum _{l=0}^{\infty} p_l \{y,x^l\}
\label{P}
\end{equation}
where 
$$ P(x) = \sum _{l=0}^{\infty} p_l x^l .$$
And we are ready to demonstrate the first two formulas which we will use 
hereafter.

{\bf Formula 1}
$$\mbox{e}^{-x} \left( y\frac{\partial }{\partial x} \right) \mbox{e}^x = 
\left\{ y,\frac{\mbox{e}^x - 1}{x} \right\}$$
This formula arises from the straightforward calculation of its left-hand-side.
We begin calculating the effect of $ y\frac{\partial }{\partial x}$ on the
left of the exponential:
$$\left( y\frac{\partial }{\partial x} \right) \mbox{e}^x = y+
\frac{1}{2}\left(yx+xy\right)+\frac{1}{3!}\left(yx^2+xyx+x^2y\right)+\ldots $$
Next we multiply on the left by $\mbox{e}^{-x}$:
$$\mbox{e}^{-x} \left( y\frac{\partial }{\partial x} \right) \mbox{e}^x = $$
$$=y+\frac{1}{2}\left(yx-xy\right)+\frac{1}{3!}\left(x^2y-2xyx+yx^2\right)+
\ldots = $$
$$=y + \frac{1}{2} \left[ y,x\right]+
\frac{1}{3!}\left[\left[y,x\right],x\right]
+\ldots=\sum _{l=0}^{\infty} \frac{1}{(l+1)!}\{y,x^l\} $$
Now,  
$$ \sum _{l=0}^{\infty} \frac{1}{(l+1)!}x^l = \frac{\mbox{e}^{x} - 1}{x},$$
so Formula 1 is demonstrated.

{\bf Formula 2}
$$\left( \left( y\frac{\partial}{\partial x}\right) \mbox{e}^{x} \right) 
\mbox{e}^{-x} = \left\{ y,\frac{1- \mbox{e}^{-x}}{x} \right\}$$
The left-hand-side of Formula 2 is obtained by multiplying the left-hand-side 
of Formula 1 on the 
left by $\mbox{e}^{x}$ and on the right by  $\mbox{e}^{-x}$. If we do the
same multiplications in the right-hand-side of Formula 1 we obtain
$$\left( \left( y\frac{\partial}{\partial x}\right) \mbox{e}^{x} \right) 
\mbox{e}^{-x} = \mbox{e}^{x}\left\{ y,\frac{\mbox{e}^{x} - 1}{x} \right\}
\mbox{e}^{-x},$$
or expanding the right-hand-side
$$\left( \left( y\frac{\partial}{\partial x}\right) \mbox{e}^{x} \right) 
\mbox{e}^{-x} =  \sum _{l=0}^{\infty} \frac{1}{(l+1)!} \mbox{e}^{x} 
\left\{ y,x^l\right\}\mbox{e}^{-x} =$$
$$ \sum _{l=0}^{\infty} \sum _{n=l}^{\infty}  \frac{(-1)^{n-l}}{(n-l)!(l+1)!}
\left\{ y,x^n\right\} = \sum _{n=0}^{\infty} c_n \left\{ y,x^n\right\}.$$
where
$$ c_n = \sum _{l=0}^n  \frac{(-1)^{n-l}}{(n-l)!(l+1)!},$$
which can be seen to correspond to the expansion of
$$\left\{ y,\frac{1- \mbox{e}^{-x}}{x} \right\}.$$
Next, Magnus demonstrates what can be called 

{\bf The Magnus' Inversion Lemma:}
{\em Let $P(x)$ and $Q(x)$ be two power series in $x$ which satisfy
$$ P(x)Q(x)=1.$$
Then each of the equations
$$ \left\{y,P(x)\right\}=u,\mbox{		}y=\left\{u,Q(x)\right\}$$
is a consequence of the other one.}
Let  $P(x)=\sum_l p_l x^l$ and $Q(x)=\sum_m q_m x^m$, then by hypothesis
$$ 1 = \sum _{l,m} p_l q_m x^{l+m}.$$
Now we can obtain the following equivalent expressions
$$ y=\left\{y,1\right\}=\left\{y, \sum _{l,m} p_l q_m x^{l+m}\right\}=
\sum _{l,m} p_l q_m \left\{y,x^{l+m}\right\},$$
where we have used notation (\ref{P}). Next we can separate indexes and 
write
$$y=\sum _m q_m\left\{\sum_l p_l\left\{y,x^l\right\},x^m\right\}=\sum _m q_m
\left\{\left\{y,P(x)\right\},x^m\right\} .$$
If we suppose now that $ \left\{y,P(x)\right\}=u $, immediately we obtain that
$$y=\sum _m q_m\left\{ u ,x^m\right\}=\left\{u,Q(x)\right\}.$$
The inverse implication is completely equivalent.

With the Inversion Lemma and Formula 2, we have all the instruments to 
solve the equation for the evolution operator

{\bf Exponential solution Theorem (Magnus):}
{\em Let ${\bf K}(t) $ be a known function of t in an associative ring (for our
purposes it is a matrix), and let ${\bf O}(t)$ be an unknown function 
(in our case the evolution operator) satisfying
\begin{equation}
\frac{d{\bf O}}{dt} =  {\bf K O} , \; {\bf O}(0)= {\bf 1},
\label{TE}
\end{equation}
where $ {\bf 1}$ is the identity matrix.
Then, if certain unspecified conditions of convergence are satisfied, 
${\bf O}(t)$  can be written in the form
$${\bf O}(t) = \exp \Omega (t)$$
where
$$\frac{d\Omega}{dt}=\left\{{\bf K},\frac{\Omega}{1-\mbox{e}^{-\Omega}}\right\}
=\sum_{n=0}^{\infty}\beta _n\left\{{\bf K},\Omega ^n\right\} = $$
$$={\bf K}+\frac{1}{2}[{\bf K},\Omega]+\frac{1}{12}\left\{{\bf K},\Omega^2
\right\}\mp \ldots $$
The $\beta _n$ vanish for $n=3,5,7,\ldots$, and 
$\beta _{2m} = (-1)^{m-1}B_{2m}/(2m)!$, where the $B_{2m}$ 
(for $m=1,2,3,\ldots$) are the Bernoulli numbers.

Integration of this equation by iteration leads to an infinite series for 
$\Omega $ the first terms of which (up to terms involving 4  ${\bf K}$'s) are
$$\Omega (t) = \int _0 ^t {\bf K}(\tau)d\tau  
+ \frac{1}{2}\int _0 ^t \left[{\bf K}(\tau), \int _0 ^\tau {\bf K}(\sigma) 
d\sigma \right] d\tau +  $$
$$ +\frac{1}{4}\int _0 ^t \left[ {\bf K}(\tau), \int _0^\tau \left[{\bf K}
(\sigma),\int _0^{\sigma} {\bf K}(\rho)d\rho \right] d\sigma \right] d\tau +$$
$$+\frac{1}{12}\int _0^t \left[ \left[ {\bf K}(\tau),\int _0^\tau {\bf K}
(\sigma)d\sigma \right],\int _0^\tau {\bf K}(\sigma)d\sigma \right] d\tau +$$
$$+\frac{1}{8}\int _0^t \left[ {\bf K}(\tau),\int _0^\tau \left[{\bf K}
(\sigma),\int _0^{\sigma}\left[ {\bf K}(\rho), \int _0^\rho {\bf K}(\nu)d\nu 
\right],d\rho \right] d\sigma \right] d\tau +$$
\begin{equation}
+\frac{1}{24}\int _0^t \left[ {\bf K}(\tau), \int _0^\tau \left\{ {\bf K}
(\sigma),\left( \int _0^{\sigma} {\bf K}(\rho)d\rho \right)^2 \right\}d\sigma 
\right] d\tau + ... 
\label{magnus}
\end{equation}}
Let us suppose ${\bf O}(t) = \exp \Omega (t)$, then
$$\frac{d{\bf O}}{dt} = \left(\left(\frac{d\Omega}{dt}
\frac{\partial}{\partial \Omega}\right)\mbox{e}^{\Omega}\right)
\mbox{e}^{-\Omega}\mbox{e}^{\Omega}.$$
By using Formula 2 with $y=\frac{d\Omega}{dt}$ and $x=\Omega$, one obtains
$$\frac{d{\bf O}}{dt} = \left\{ \frac{d\Omega}{dt},\frac{1-\mbox{e}^{-\Omega}}
{\Omega}\right\}\mbox{e}^{\Omega},$$
which compared with Eq.(\ref{TE}) gives
$${\bf K} (t) =\left\{ \frac{d\Omega}{dt},\frac{1-\mbox{e}^{-\Omega}}
{\Omega}\right\} .$$
We can now apply the Inversion Lemma with the same substitutions in $x$ and $y$
as before, and with $u={\bf K}$ and 
$$P(\Omega )=\frac{1-\mbox{e}^{-\Omega}}{\Omega},$$
to obtain 
$$Q(\Omega )=\frac{\Omega}{1-\mbox{e}^{-\Omega}}$$
and
$$\frac{d\Omega}{dt}=\left\{{\bf K},\frac{\Omega}{1-\mbox{e}^{-\Omega}}\right\}
.$$
This expression can be finally expanded using the following power series
$$\frac{\Omega}{1-\mbox{e}^{-\Omega}}=\sum_{n=0}^{\infty}\beta _n \Omega ^n$$
where the $\beta _n$ has the given values.

To integrate the resulting equation for $\Omega$ we start at t=0, where
$\Omega (0) = 0$ to satisfy boundary conditions. Introducing this solution
into the equation we obtain a new solution:
$$ \Omega _1 = \int _{0}^{t}{\bf K} d\tau .$$
We can iterate the procedure to obtain $ \Omega _2$, $ \Omega _3$, \ldots as
$$ \frac{d\Omega _m}{dt}=\sum_{n=0}^{\infty}\beta _n \left\{{\bf K},
\Omega_{m-1} ^n\right\}.$$
The solution is obtained as the limit of this series:
$$ \Omega(t) = \lim _{m \rightarrow \infty}  \Omega _m .$$

\begin{acknowledgements}
We gratefully thank J.C. del Toro Iniesta, A. Skumanich and G. Chambe for 
carefully reading this paper  and for their valuable comments.
\end{acknowledgements}



\end{document}